\documentclass[12pt]{article}
\usepackage{graphicx}

\def\jlab{Jefferson Laboratory\\
12000 Jefferson Avenue, Newport News, Virginia 23606, USA}
\def\yerphi{Yerevan Physics Institute, Yerevan, Armenia}
\def\Title#1{\begin{center} {\Large #1 } \end{center}}
\def\Author#1{\begin{center}{ \sc #1} \end{center}}
\def\Address#1{\begin{center}{ \it #1} \end{center}}

\newenvironment{Abstract}{\begin{quotation}  }{\end{quotation}}
\newenvironment{Presented}{\begin{quotation} \begin{center} 
             PRESENTED AT\end{center}\bigskip 
      \begin{center}\begin{large}}{\end{large}\end{center} \end{quotation}}
\def\Acknowledgements{\bigskip  \bigskip \begin{center} \begin{large}
             \bf ACKNOWLEDGEMENTS \end{large}\end{center}}

\begin{document}
\begin{titlepage}

\Title{\huge Nucleon Resonance Transition Formfactors \\ }
\vfill
\Author{Volker D. Burkert, Victor I. Mokeev }
\Address{\jlab}
\Author{I. G. Aznauryan} 
\Address{\yerphi}
\vfill
\date{today}
\begin{Abstract}
We discuss recent results from CLAS on electromagnetic resonance transition amplitudes and their 
dependence on the distance scale ($Q^2$). 
From the comparison of these results with most advanced theoretical
calculations within QCD-based approaches there is clear evidence
that meson-baryon contributions are present and important at large distances, i.e. small $Q^2$, 
and that quark core contributions dominate the short distance behavior.
\end{Abstract}
\vfill
\begin{Presented}
Conference on Intersections of Particle and Nuclear Physics,\\ CIPANP 2015, Vail, CO, May 19-24, 2015 
\end{Presented}
\vfill
\end{titlepage}
\def\thefootnote{\fnsymbol{footnote}}
\setcounter{footnote}{0}

\section{Introduction}
The excited states of the nucleon have been studied experimentally 
since the development of the quark model in 1964 \cite{GellMann1964,Zweig1964}.
The 3-quark structure of the baryons  when realized in the dynamical 
quark models resulted in prediction of a wealth of excited states  
with underlying spin-flavor and orbital symmetry of $SU(6) \otimes O(3)$. Most of 
the initially observed states were found with hadronic probes.  From the many excited 
states predicted by the quark model, only a fraction have been observed to date.  
The search for the "missing" states  and detailed studies of the resonance structure  
is now mostly carried out using electromagnetic probes and has been a major focus of  
hadron physics for the past decade \cite{Burkert:2004sk}. This has led to a broad 
experimental effort and the measurement
of  exclusive meson photoproduction and electroproduction reactions, including 
many polarization observables. As a result, several new excited states of the 
nucleon have been discovered 
and entered in the Review of Particle Physics~\cite{Agashe:2014kda}. 

Meson electroproduction, which is the subject of this talk, has revealed intriguing new 
information regarding the relevant degrees of freedom underlying  the structure of the 
excited states at different distance scale probed \cite{Aznauryan:2011qj}.  

\section{The $\rm N\Delta(1232)$ transition}

One of the important insights is clear evidence that resonances are not excited 
from quark transitions alone, but there can be significant contributions from 
meson-baryon interactions as well, and that these two processes contribute to the 
excitation of the same state. This evidence has been obtained in part through 
the observation that the 
quark transition processes  often do not have sufficient strength to explain fully 
the measured resonance transition amplitudes. The best known example is the 
$\Delta(1232){3/2}^+$ resonance, which, when excited electromagnetically is mostly 
due to a magnetic dipole transition from the nucleon ground state, but only a fraction 
of the magnetic transition form factor can be explained by the quark content of the state
at the photon point. Instead, at $Q^2 \ge 3$~GeV$^2$ the quark content becomes the biggest 
contributor to this transition form factor. 
At low $Q^2$ a satisfactory description of this transition can be achieved in models that include 
pion-cloud contributions  
and also in dynamical reaction models, where the missing strength has been attributed 
to dynamical meson-baryon interaction in the final state~\cite{Sato2008}. 

\begin{figure}[t]
\centering
\includegraphics[height=2.50in]{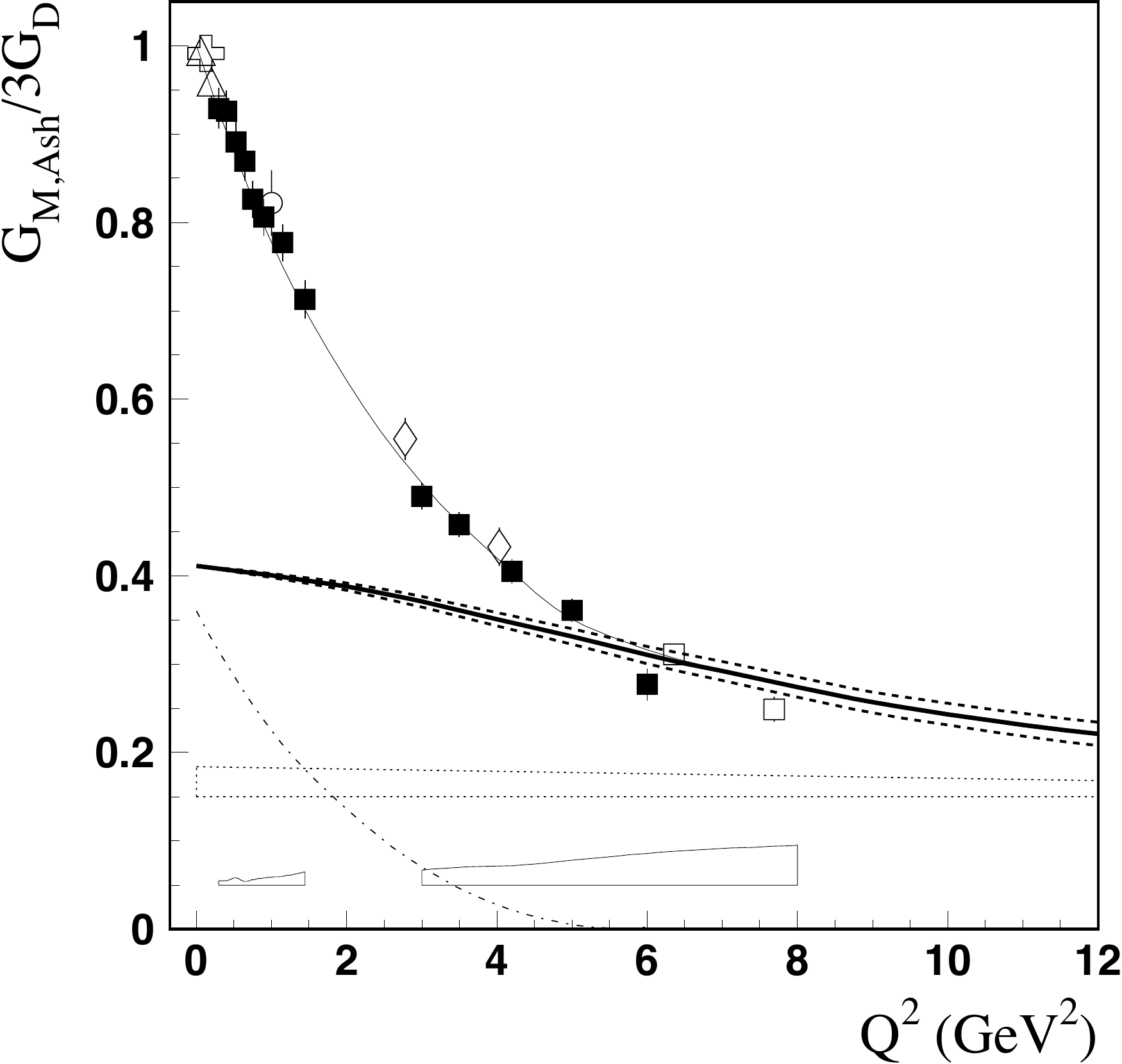}
\caption{\small The magnetic transition form factor $G_M^\ast$ of the $N\Delta(1232)$ transition as
determined in various experiments and analyses of $ep \to ep\pi^0$. The thin solid line is the 
global analysis of Ref.~\cite{Drechsel:2007if}.  The 3-quark 
contributions in the LF RQM (solid line with dotted error band) is normalized to the high $Q^2$ behavior.  
The open band at the bottom shows the model uncertainties of the extracted data points. The band above it
represents the uncertainties of the model analysis~\cite{Aznauryan:2015zta}. The thin dash-dotted line is the 
estimated meson-baryon contribution from Ref.~\cite{Sato2001}.}
\label{fig:Delta}
\end{figure}

A recent calculation within the light front relativistic quark model 
(LF RQM)~\cite{Aznauryan:2012ec,Aznauryan:2015zta} with the 3-quark contributions that are 
 normalized to the high $Q^2$ behavior finds that 
at the photon point more than 50\% of the strength may be due to non-quark contributions, 
as shown in Fig.~\ref{fig:Delta}.    
The excitation of this and other states using electron beam should be highly sensitive to the different Fock 
components in the wave function of the excited states as it is expected that they have different
excitation strengths when probed at large and at short distance scales, 
i.e. with virtual photons at low and high $Q^2$. At high $Q^2$ we expect 
the $qqq$ components to be the only surviving part, while the higher Fock states may have large, 
even dominant strength at low $Q^2$.     
\section{Solving the Roper puzzle}
It is known that the Roper $N(1440){1\over 2}^+$ state presented the 
biggest puzzle of the well-established resonances and defied explanations within the quark model 
for decades. The constituent quark model has it as the first radial excitation of the 
nucleon ground state. However, its physical mass is about 300 MeV lower than what is predicted. 
The most recent LQCD projections have the state even 1 GeV above the nucleon ground state, 
i.e. near 1.95GeV~\cite{Dudek:2012ag}. 
The electromagnetic transition amplitude extracted from pion photo production data is large and negative, while 
the non-relativistc constituent quark model (nrCQM) predicts a large and positive amplitude. 
Furthermore, the early electroproduction results showed a rapid disappearance of its  
excitation strength at $Q^2 \leq 0.5$GeV$^2$, while the model predicted a strong rise in 
magnitude. These apparent discrepancies led to attempts at alternate interpretations of the state, 
e.g. as the lowest gluonic excitation of the nucleon~\cite{Barnes:1982fj}, 
and as dominantly $N\rho$~\cite{Cano:1998wz} or $N\sigma$~\cite{Faessler} molecules.    
Recent development of the dynamically coupled channel models by the EBAC group, has  
led to a possible resolution of the discrepancy in the mass values, by including resonance coupling to inelastic 
decay channels in their calculations~\cite{Suzuki:2009nj}. The inelastic channels caused the dressed Roper pole to move 
by over 350 MeV close to 1.365~GeV from the bare value of 1.736 GeV,  i.e. close to 
whereit is found experimentally. 
\begin{figure}[t]
\hspace{-0.4cm}
\includegraphics[height=4.in,width=5.50in]{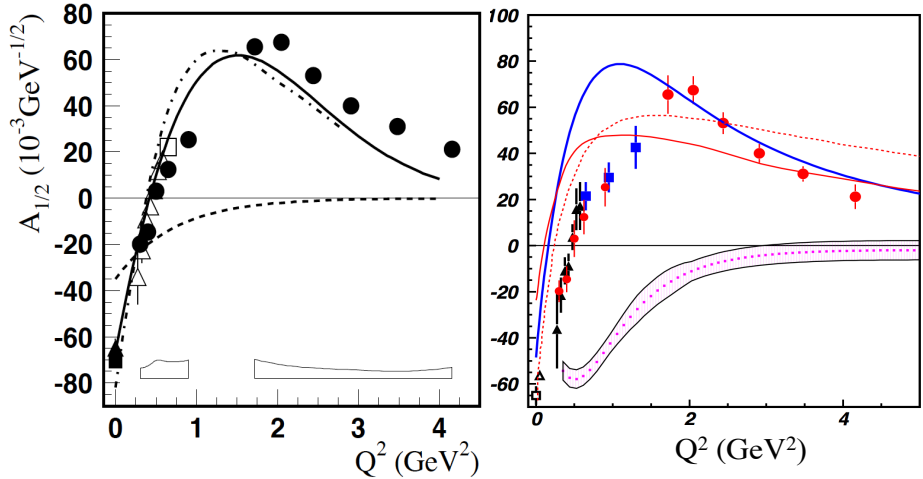}
\vspace{-2cm}
\caption{\small The transverse electrocoupling amplitude $A_{1/2}(Q^2)$ for the Roper $N(1440){1\over 2}^+$. The panel 
on the left shows the CLAS data as of 2012~\cite{Aznauryan:2009mx,Mokeev:2012vsa}. The curves are non-relativistic quark model 
calculations that include $N\sigma$ configurations~\cite{Obukhovsky:2011sc}. The dashed line is the $N\sigma$ contribution, 
the solid line is the full result. At high $Q^2$  the model 
undershoots the data significantly. The right panel includes also recent data from $p\pi^+\pi^-$ final 
states~\cite{Mokeev:2015lda} (blue squares). 
The red curves are LF RQM calculations with $N\sigma$ components~\cite{Obukhovsky:2013fpa} (dashed), and with 
momentum-dependent quark masses~\cite{Aznauryan:2012ec} (solid).  The blue curve is the DSE/QCD 
calculation~\cite{Segovia:2015hra} after renormalization of the quark wave function~\cite{Mokeev:2015lda}. 
The band in the lower part represents the difference of the DSE/QCD predictions and the CLAS data. }
\label{fig:a12roper}
\end{figure}
Measurements of the electro-coupling amplitudes in large range of $Q^2$~\cite{Park:2007tn,Fedotov:2008aa,Aznauryan:2009mx} 
provided strong evidence for the Roper resonance as a predominantly first radial excitation of a nucleon ground state.The electrocoupling 
amplitudes are shown in Fig.~\ref{fig:a12roper}.  
The LF RQM predict correct sign of the transverse amplitude at $Q^2=0$ and a sign cross-over at small $Q^2$. The 
behavior at low $Q^2$ is described well when the 3q component in the wave function is complemented by 
meson-baryon contributions, e.g. $N\rho$~\cite{Cano:1998wz} and  $N\sigma$~\cite{Obukhovsky:2011sc}, 
and also in effective field theories~\cite{Bauer:2014cqa} employing pions, $\rho$ mesons, 
the nucleon and the Roper $N(1440){1\over 2}^+$ as effective degrees of freedom. 
The high $Q^2$ behavior is well reproduced in the QCD/DSE approach and the LF RQM which include 
momentum-dependent quark masses, in QCD/DSE~\cite{Segovia:2015hra} due to full incorporation 
of the momentum-dependent dressed quark mass in QCD and in LF RQM~\cite{Aznauryan:2012ec} 
by a parameterized mass function.   

\section{The parity partner of the nucleon $N(1535){1\over 2}^-$}
\begin{figure}[t]
\hspace{1.0cm}\includegraphics[height=2.0in,width=4.5in]{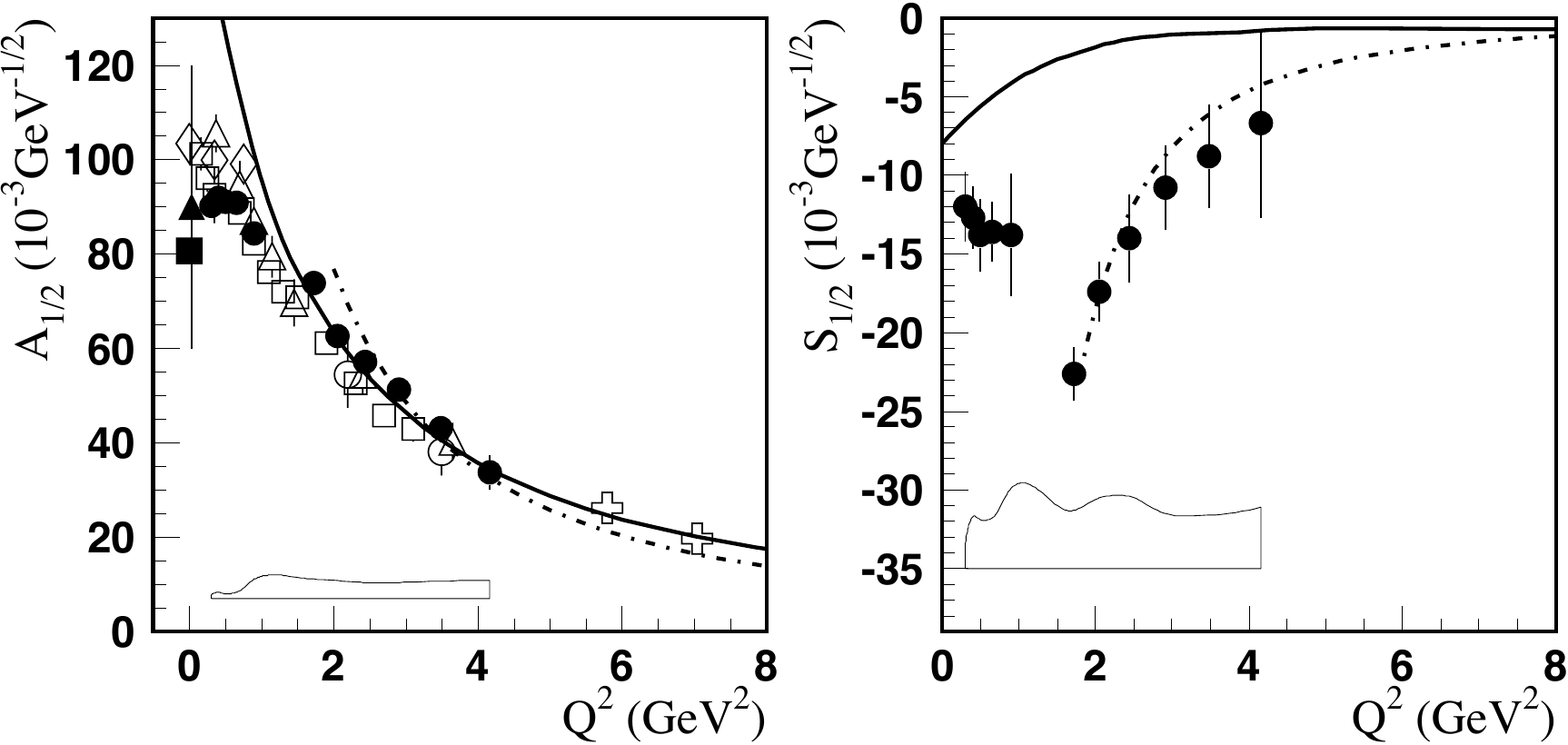}
\caption{\small The transverse and scalar amplitudes for the $N(1535){1\over 2}^-$  determined in 
$ep \to ep\eta$ (open symbols) and in $ep \to eN\pi$ (full circles). 
The open bands indicate the size of model-dependency in the extraction 
of the resonance amplitudes. Curves represent LF RQM (solid), 
and LC SR (dashed-dotted). }
\label{fig:N1535}
\end{figure}

The parity partner of the ground state nucleon lies 600 MeV above the mass of the nucleon. The shift 
is thought to be due to the breaking of chiral symmetry in the excitation of nucleon resonances. 
The state has been difficult to interpret in terms quark excitations only, especially the sign and 
$Q^2$-dependence of its scalar amplitude. Figure~\ref{fig:N1535} shows both  
$A_{1\over 2}$ and $S_{1\over 2}$ transition amplitudes. $A_{1\over 2}$ is well described by the 
LF RQM~\cite{Aznauryan:2012ec} and 
the  LC SR (NLO)~\cite{Anikin:2015ita} evaluation for $Q^2 \geq 1.5 $GeV$^2$. 
The scalar amplitude $S_{1\over 2}$ 
departs from the LF RQM predictions significantly, it is, however, well described by the LC SR (NLO) 
calculation at $Q^2 \ge 1.5$GeV$^2$.  The lowest moments of the $N(1535){1\over 2}^-$ quark distribution 
amplitudes fit to the electrocoupling data are consistent with the values computed in LQCD~\cite{Anikin:2015ita}.
This result points at a promising approach of relating the resonance electrocouplings to calculations from 
first principle of QCD.  
The state has also been discussed as having large strangeness components~\cite{Zou:2006tw}, 
an assertion that might account for the discrepancy in the scalar amplitude with the data at low $Q^2$.  

\section{Transverse LF charge transition densities.} 

\begin{figure}[t]
\vspace{-2.0cm}
\hspace{1.0cm}
\includegraphics[height=4.0in,width=5.0in]{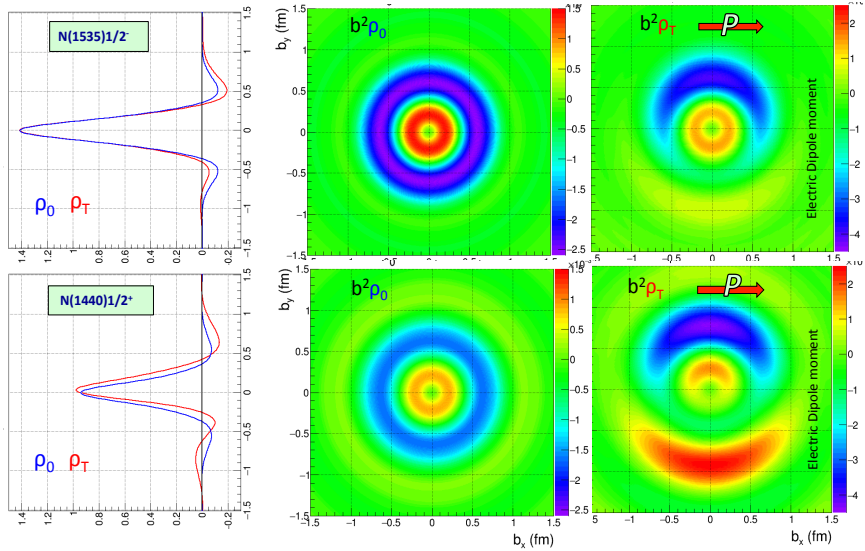}
\vspace{-1.5cm}
\caption{\small The transverse charge transition densities of the $N(1440){1\over 2}^+$ and $N(1535){1\over 2}^-$ 
using the definition of $\rho_0$ and $\rho_T$ in Ref.~\cite{Tiator:2008kd}. 
For comparison of the two states, the same scales have been used in all graphs. The two left 
panels show the charge densities projected on the b$_y$ axis; $\rho_0$ (blue line) is for unpolarized protons, $\rho_T$ 
for transversely polarized protons. The 2-D plots on the right show the b$_x$ versus b$_y$ correlations of the $\rho_0$ (middle) and
$\rho_T$ (right) transition densities.
To emphasize the large distance behavior the densities have been scaled with b$^2$ causing the hole in the center. }
\label{fig:charge_densities}
\end{figure}

Knowledge of the electrocoupling amplitudes in a large range of $Q^2$ allows for the determination of the transition 
charge densities in the light front~\cite{Tiator:2008kd}. Such densities have been obtained for the 
$\Delta(1232){3\over 2}^+$ where data are available from the photon point to $Q^2 \approx 7.0$GeV$^2$ for 
all 3 electrocoupling amplitudes and for 
states in the second nucleon resonance region. Fig.~\ref{fig:charge_densities} shows the charge transition 
densities based on new fits 
to the $A_{1/2}(Q^2)$ and $S_{1/2}(Q^2)$ of the $N(1440){1\over 2}^+$ and $N(1535){1\over 2}^-$ electrocoupling amplitudes. 
This allowed the extraction of transition charge densities for the unpolarized 
nucleon to the excited state, and for the transversely polarized proton to the excited states. One would expect that
the charge densities are different for the two states as one is a radial excitation of the nucleon, the other an orbital excitation
of the quark core. There is indeed a  
notable difference between the two states. The Roper $N(1440)$ shows a considerably softer core and wider wings compared to
the $N(1535)$. Furthermore, the peak in the $N(1440)$ charge distribution for the polarized proton moves away from the center to more 
positive b$_y$ generating a strong electric dipole along b$_y$, while the $N(1535)$ core shows no change in position.  

\section{New results on states in the 1.7 GeV mass range}
\begin{figure}[t]
\vspace{-2.5cm}
\includegraphics[height=4.0in,width=5.5in]{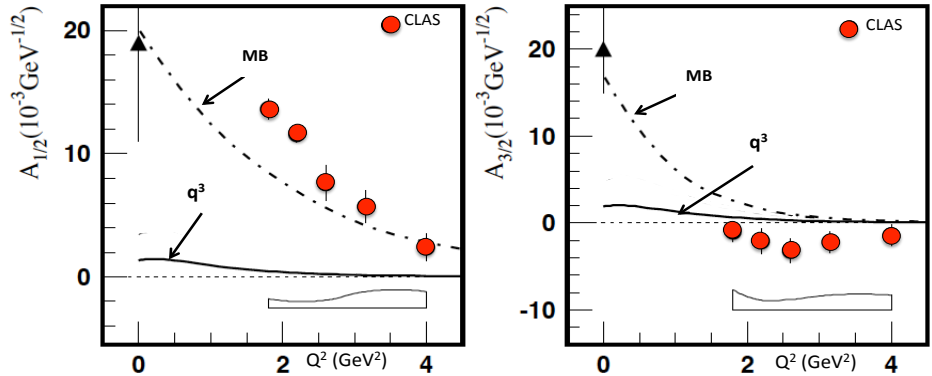}
\vspace{-4.0cm}
\caption{\small The transverse amplitudes of the $N(1675){5\over 2}^-$ have been determined in $ep \to e\pi^+n$
(for references see text).}
\label{fig:N1675}
\end{figure}

\begin{figure}[t]
\includegraphics[height=2.0in,width=5.5in]{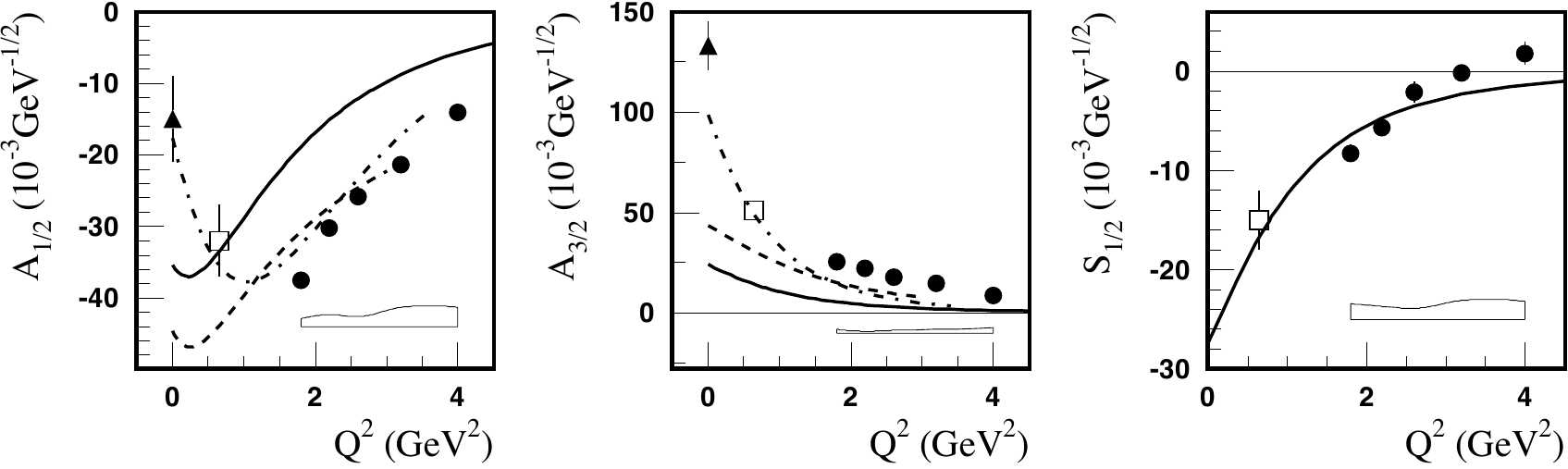}
\caption{\small The transverse and scalar amplitudes for the $N(1680){5\over 2}^+$ determined in $ep \to e\pi^+n$. 
The curves are projections of various quark models.}
\label{fig:N1680}
\end{figure}

Cross sections on $ep\to e\pi^+n$ have been published recently in the mass range from 1.6 to 2.0 GeV~\cite{Park:2014yea}. 
The so-called third resonance regions is the domain of several nearly mass degenerate states with masses near 1.7~GeV. 
Several states, e.g. $N(1675){5\over 2}^-$ and $N(1650){1\over 2}^-$ belong in $SU(6)\otimes O(3)$ to 
the $[70,1^-]_1$ supermultiplet, while the $N(1680){5\over 2}^+$ quark state is assigned to $[56, 2^+]_2$.
The assignment to different multiplets within that symmetry group has important impact on the transition 
strength of the 3-quark components in the excited states. Depending on the multiplet assignment, we may 
expect quite different strengths and $Q^2$ dependences of the quark components. 
For example, the quark structure of the $N(1675){5\over 2}^-$ leads to a suppressed 3-quark transition amplitude 
from the proton, i.e. $A^q_{1/2} = A^q_{3/2} = 0$. We employed this suppression 
to directly access the non-quark components~\cite{Aznauryan:2014xea}. The results for the $N(1675){5\over 2}^-$ 
 are shown in  Fig.~\ref{fig:N1675}. The constituent quark model predictions are from 
Ref.~\cite{Santopinto:2012nq}. Shown predictions for the meson-baryon (MB) contributions are absolute values of 
the results from the dynamical coupled-channel model (DCCM)~\cite{JuliaDiaz:2007fa}.They are in qualitative 
agreement with the amplitudes extracted from experimental data, i.e. considerable coupling through the $A_{1/2}$ 
amplitude and much smaller $A_{3/2}$ amplitude at $Q^2 \ge 1.8$GeV$^2$. Figure~\ref{fig:N1680} shows the results for 
the $N(1680){5\over 2}^+$ resonant state.  There is  a rapid drop with $Q^2$ of the $A_{3/2}$ amplitude, which 
dominates at $Q^2=0$, while the $A_{1/2}$ amplitude, which at $Q^2=0$ makes a minor contribution,  
becomes the leading amplitude at larger $Q^2$. This change of the helicity structure is expected, 
but it is less rapid than predicted by quark models, which could hint at sizable meson-baryon contributions at  
large $Q^2$. 
\section{Conclusions}
The meson electroproduction program at Jefferson Lab has revealed that many excited states predicted within the 
symmetry group $SU(6)\otimes O(3)$ have higher Fock state components that can be represented by 
meson-baryon components in the wave function. Most states in the mass range up to 1.7 GeV exhibit the common 
feature that the excitation strengths of the higher Fock states components decrease more rapidly with increasing $Q^2$ 
than the leading qqq components. At $Q^2 \ge 2-3$GeV$^2$ the qqq components dominate and closely follow the LF RQM 
calculations~\cite{Aznauryan:2012ec} and the projections of DSE/QCD~\cite{Segovia:2015hra} and NLO LCSR~\cite{Anikin:2015ita}. 
The results on the Roper resonance $N(1440){1\over 2}^+$ clearly establish the state as the first radial excitation of the 
nucleon's 3-quark core. The most recent theoretical developments successfully explored new avenues towards relating resonance 
electrocouplings to the first principles of QCD. 
\Acknowledgements
We thank F.X. Girod for preparing the graphs on the transition charge densities. This work was supported by the US Department of Energy, Office of Science, Office of Nuclear Science, under contractDE-AC05-06OR23177,  and by the State Committee of Science of Republic of Armenia under Grant 13-1C023.

\end{document}